\def\eV{{\,\textrm{eV}}}

\def\aa{{\,\textrm{\AA}}}

\documentclass[twocolumn,superscriptaddress,preprintnumbers,amsmath,amssymb,prb]{revtex4}

\usepackage{graphicx}
\usepackage{epstopdf}
\usepackage{setspace}

\usepackage{subfigure}
\usepackage{dcolumn}
\usepackage{bm}
\usepackage{color}

\begin{document}

\title{ \bf{All-Electron GW Quasiparticle Band Structures of Group 14 Nitride Compounds}}

\author{Iek-Heng Chu} 
\affiliation{Department of Physics and Quantum Theory Project, University of Florida, Gainesville, Florida 32611, United States}
\author{Anton Kozhenikov}
\affiliation{Institute for Theoretical Physics, ETH Zurich, 8093 Zurich, Switzerland}
\author{Thomas C. Schulthess}
\affiliation{Institute for Theoretical Physics, ETH Zurich, 8093 Zurich, Switzerland}
\author{Hai-Ping Cheng}
\email[Corresponding author: Hai-Ping Cheng,  ]{Email: cheng@qtp.ufl.edu}
\affiliation{Department of Physics and Quantum Theory Project, University of Florida, Gainesville, Florida 32611, United States}

\begin{abstract}
{We have investigated the group 14 nitrides (M$_3$N$_4$) in the spinel phase ($\gamma$-M$_3$N$_4$ with M= C, Si, Ge and Sn) and $\beta$ phase ($\beta$-M$_3$N$_4$ with M= Si, Ge and Sn) using density functional theory with the local density approximation and the GW approximation. The Kohn-Sham energies of these systems have been first calculated within the framework of full-potential linearized augmented plane waves and then corrected using single-shot G$_0$W$_0$ calculations, which we have implemented in the modified version of the Elk full-potential LAPW code. Direct band gaps at the $\Gamma$ point have been found for spinel-type nitrides $\gamma$-M$_3$N$_4$ with M= Si, Ge and Sn. The corresponding GW-corrected  band gaps agree with experiment. We have also found that the GW calculations with and without the plasmon-pole approximation give very similar results, even when the system contains semi-core $d$ electrons. These spinel-type nitrides are novel materials for potential optoelectronic applications because of their direct and tunable band gaps. } 
\end{abstract}

\maketitle

\section{Introduction}
Group 14 nitride compounds, M$_3$N$_4$ (M=C, Si, Ge and Sn), are an important class of semiconductors. Among them is silicon nitride (Si$_3$N$_4$), which has been extensively studied by theoretical and experimental groups\cite{PhysRevB.61.8696, PhysRevB.55.3456, Bradley19661803, PhysRevB.84.045308, PhysRevB.41.10727, bsi, bge}. It is known that Si$_3$N$_4$ can exist in two energetically favorable phases, $\alpha$- and $\beta$-Si$_3$N$_4$, which have hexagonal crystal structures with different stacking patterns of the layered atoms perpendicular to the $c$ axis. Since the first discovery of Si$_3$N$_4$ in a third phase, the cubic spinel phase\cite{Zerrsi3n4} ($\gamma$-Si$_3$N$_4$) reported in 1999, this class has stimulated great research efforts in the past few years\cite{Yang_ge3n4, PhysRevB.61.11979, JACE75, PhysRevB.73.045202, PhysRevLett.83.5046, PhysRevB.78.064104, PhysRevLett.111.097402, PhysRevB.82.144112, Huang_sn3n4}. $\gamma$-Ge$_3$N$_4$ and $\gamma$-Sn$_3$N$_4$ have also been synthesized successfully in subsequent experiments\cite{r-sn3, r-ge3, r-ge32}, however $\gamma$-C$_3$N$_4$ has not yet been found experimentally. Whereas $\gamma$-Si$_3$N$_4$ and $\gamma$-Ge$_3$N$_4$ are synthesized in a high-temperature and high-pressure environment, the synthesis of $\gamma$-Sn$_3$N$_4$ can be achieved at ambient conditions. These novel spinel compounds exbihit remarkable mechanical properties and high thermal stability, and they form a new class of superhard materials. Also, with interesting electronic properties such as direct and tunable band gaps, these materials are good candidates for optoelectronic applications, e.g. light-emitting diodes. 

Previous \textit{ab initio} electronic structure calculations were carried out for these nitride compounds in different phases\cite{PhysRevB.61.8696, Yang_ge3n4, PhysRevB.61.11979, JACE75, PhysRevB.63.064102, c3n4_op}. The methods used were mainly based on density functional theory (DFT)\cite{KSE}, either within the orthogonalized linear combinations of atomic orbitals (OLCAO) method\cite{olcao} or the plane-wave pseudopotential method. The studies concerning electronic band gaps are important for a variety of potential applications. However, it is well known that DFT is a ground-state theory and it severely underestimates the band gap, which relates to the functional derivative discontinuity of the exchange-correlation potential. The DFT Kohn-Sham (KS) energies may also not match the quasiparticle energies because of a lack of many-body interactions. The standard calculation including many-body corrections uses the GW approximation proposed by Hedin\cite{gw1} within the framework of many-body perturbation theory. Recently, Kresse \textit{et al.}\cite{PhysRevB.85.045205} have studied the GW-corrected electronic properties of $\alpha$- and $\beta$-Si$_3$N$_4$. Gao \textit{et al.}\cite{Gao2013292} have applied the GW approach to study the band gaps for $\alpha$-, $\beta$- and $\gamma$-Ge$_3$N$_4$. Xu \textit{et al.}\cite{Xu201211072} have studied the GW quasiparticle energies of the C$_3$N$_4$ polymorphs. All these calculations have been performed using the plane-wave based pseudopotential or projected augmented wave (PAW)\cite{PhysRevB.50.17953} method. In current practice, neither pseudopotential nor PAW methods includes GW for core level corrections and there is evidence that the GW results calculated using the pseudo wave functions might lead to errors compared with those using the all-electron wave functions\cite{PRL.101.106404,AE_GW}. 

Another common approximation uses the fact that the inverse dielectric matrix is usually peaked around the plasma frequency but very flat elsewhere.  This plasmon-pole approximation (PPA) is often introduced to simplify the frequency dependence of that matrix and to speed up the GW calculations. The PPA has been found to yield band gaps close to experimental results. However, it is also known that the results from different PPA models\cite{gw2,GNPPM,vdLHPPM,EFPPM} may vary substantially when compared with those from the numerical integration (NI) method, due to the different parameter-fitting conditions.  Recent studies show that the PPA model proposed by Godby and Needs (GN)\cite{GNPPM} agrees consistently with the NI method\cite{PhysRevLett.100.186401, PhysRevB.84.241201, PhysRevB.88.125205}, however it has been discussed that the PPA can become questionable when it is applied to systems with localized electrons\cite{gw_review}, e.g. semi-core $d$ electrons. A careful study on this issue is therefore needed.

In this work, we have investigated the electronic structures of the $\gamma$-M$_3$N$_4$ (M=C, Si, Ge and Sn). The DFT KS eigenvalues and eigenfunctions have been computed within the framework of the full-potential linearized augmented plane wave (LAPW) method, which does not require pseudopotentials. We have then used the KS energies as input for the single-shot G$_0$W$_0$ to compute the quasiparticle energies. We have also studied the $\beta$-M$_3$N$_4$ (M= Si, Ge and Sn), in which the $\beta$ phase is found to be more energetically stable for Si$_3$N$_4$ and Ge$_3$N$_4$. In the GW calculations, both the GN-PPA model and the NI method have been used and results have been compared. The rest of the article is organized as follows: we give a brief overview of the GW formalism including details of our implementation in Section II, describe our model systems and the computational details in Section III, present our results in Section IV, and end with conclusions in Section V.

\section{Methodology}

\subsection{The GW method}

Within the single-particle picture, the quasiparticle equation reads
\begin{eqnarray}
&&[T+V_{ext}({\bf r})+V_{H}({\bf r})]\Psi_{nk}({\bf r})+\label{QPE}\\\nonumber
&&\int d{\bf r'}~\Sigma({\bf r},{\bf r'},E^{QP}_{nk})\Psi_{nk}({\bf r})=E^{QP}_{nk}\Psi_{nk}({\bf r}),
\end{eqnarray}
where $T$, $V_{ext}$ and $V_{H}$ are the kinetic energy operator, the external potential from the nuclei and the Hartree potential, respectively. $\Sigma$ is the self-energy operator that accounts for all the many-body electron-electron interactions beyond the Hartree term. In general, the self-energy is energy-dependent, non-local and non-Hermitian. This leads to complex quasiparticle energies $E_{nk}$ of which the imaginary part relates to the quasiparticle lifetime. In Eq.(\ref{QPE}), since the self-energy depends on $E^{QP}_{nk}$, this equation has to be solved self-consistently, which is computationally very expensive.

The practical method for calculating the self-energy employs the GW approximation\cite{gw1}, in which the self-energy is expanded and only the first-order term is retained.  When expressed in real space, the self-energy within this approximation reads as
\begin{equation}
\Sigma({\bf r},{\bf r'},\omega)=\frac{i}{2\pi}\int d\omega'~G({\bf r},{\bf r'},\omega+\omega')W({\bf r},{\bf r'},\omega')e^{i\omega'\eta},
\end{equation}
where $G({\bf r},{\bf r'},\omega)$ is the single-particle Green function, $\eta$ is a positive infinitesimal and $W({\bf r}.{\bf r'},\omega)=\int d{\bf r_1}\varepsilon^{-1}({\bf r},{\bf r}_1,\omega)v(|{\bf r}_1-{\bf r'}|)$ is the dynamically screened Coulomb potential, with $\varepsilon$ and $v$ being the dielectric function and the bare Coulomb potential. The dielectric function can be constructed using the density-density response function in the random phase approximation (RPA), which is detailed in the next sub-section.

In conventional GW calculations, the quasiparticle eigenfunctions are approximated as the DFT Kohn-Sham (KS) eigenfunctions. This is based on previous observations that the quasiparticle eigenfunctions change little compared with the KS eigenfunctions. The self-energy matrix in the KS basis becomes diagonal, i.e. only the diagonal elements of the self-energy matrix need to be evaluated. For single-shot G$_0$W$_0$, in which one non-self consistent GW loop is performed, the KS eigenvalues $\epsilon_{nk}$ and eigenfunctions $\Psi_{nk}({\bf r})$ are used to construct the non-interacting Green function $G_0$ and the screened potential $W_0$, and hence the self-energy. The quasiparticle energies can then be calculated using first-order perturbation theory:
\begin{eqnarray}
&& E^{QP}_{n{\bf k}}=\epsilon_{n{\bf k}}+Z_{n{\bf k}}[\Sigma_{n{\bf k}}(\epsilon_{n{\bf k}})-V^{xc}_{n{\bf k}}],\label{eqp}\\
&& Z_{n{\bf k}}=[1-\frac{d\Sigma_{n{\bf k}}(\omega)}{d\omega}|_{\omega=\epsilon_{n{\bf k}}}]^{-1}.
\end{eqnarray}
Here, $\Sigma_{n{\bf k}}(\omega)\equiv\langle\Psi_{n{\bf k}}|\Sigma(\omega)|\Psi_{n{\bf k}}\rangle$, $Z_{n{\bf k}}$ is known as the renormalization factor for the KS state with band index $n$ and wave vector ${\bf k}$. $V^{xc}$ is the KS exchange-correlation potential.

\begin{figure}
\includegraphics[width=7.5cm]{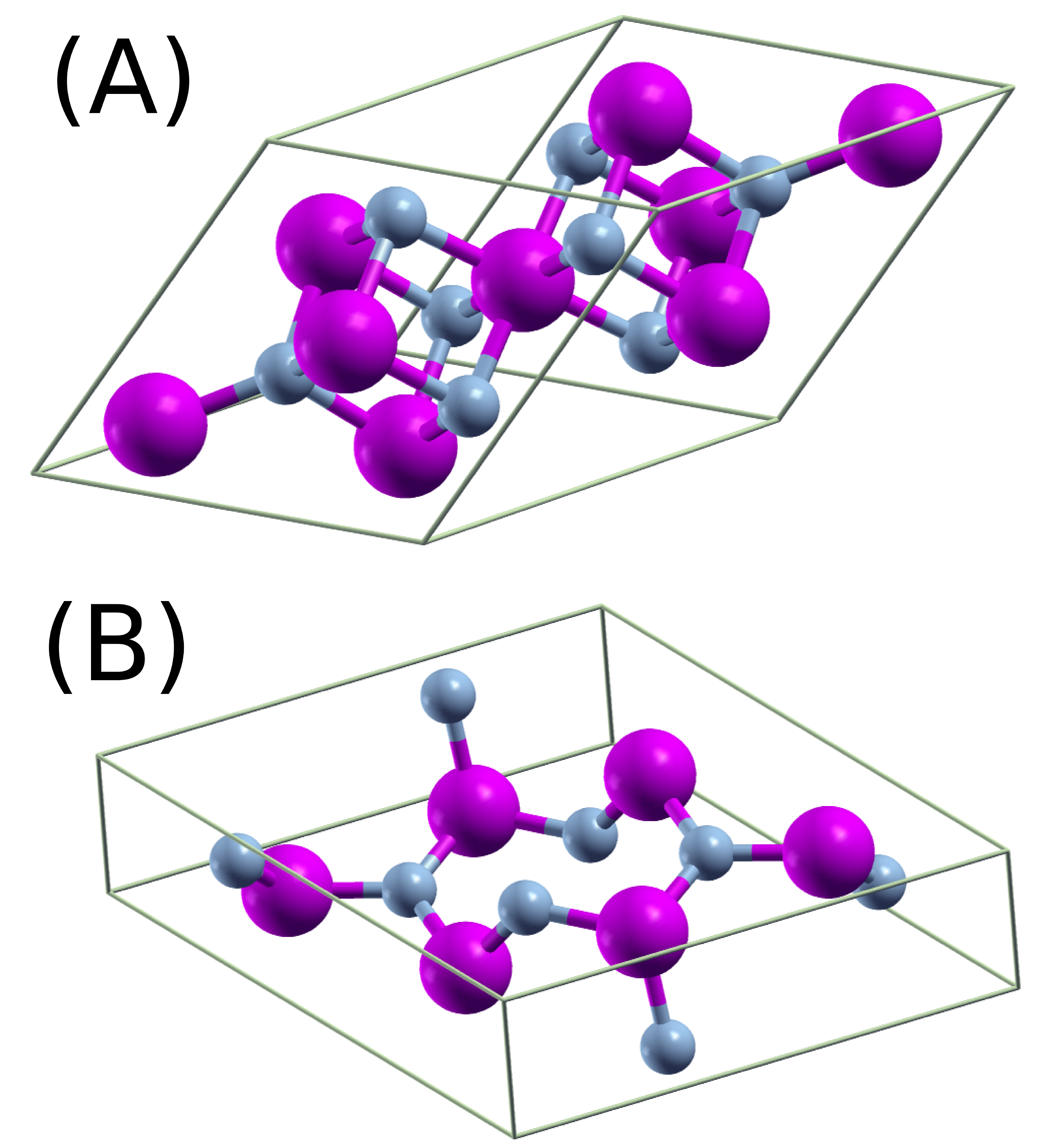}
\caption{(color online) The primitive cell of the group 14 nitride compounds in (A) spinel-phase $\gamma$-M$_3$N$_4$ (face-centered cubic) and (B) $\beta$-phase $\beta$-M$_3$N$_4$ (hexagonal). Purple and grey balls refer to M (M=C, Si, Ge and Sn) and nitrogen atoms, respectively.}\label{cell}
\end{figure}

\subsection{Evaluation of the self-energy}
We have implemented the G$_{0}$W$_{0}$ approach in the modified version of the Elk full-potential LAPW  code\cite{elk}. To calculate the self-energy matrix in the KS basis, we have split it into two terms: $\Sigma_{n{\bf k}}(\omega)=\Sigma^x_{n{\bf k}}+\Sigma^c_{n{\bf k}}(\omega)$ where $\Sigma^x$ is the exchange self-energy matrix whose elements are known as the Fock terms in the Hartree-Fock approximation. $\Sigma^c$ is the correlation self-energy matrix. $\Sigma^x$ within the KS basis, $\Sigma^x_{n{\bf k}}\equiv\langle\Psi_{n{\bf k}}|iG(\omega)v|\Psi_{n{\bf k}}\rangle$ can be written as
\begin{eqnarray}
 \Sigma^x_{n{\bf k}}&=&-\sum\limits_{{\bf k}',\sigma}\sum^{occ}_{m}\int d{\bf r}d{\bf r'}\frac{[\psi^{\sigma}_{n{\bf k}}({\bf r})]^*\psi^{\sigma}_{m{\bf k}'}({\bf r})}{|{\bf r}-{\bf r'}|}\\\nonumber
&&\times[\psi^{\sigma}_{m{\bf k}'}({\bf r'})]^*\psi^{\sigma}_{n{\bf k}}({\bf r'}),
\end{eqnarray}
where the KS eigenfunctions \{$\Psi_{n{\bf k}}({\bf r})$\} are expressed in spinor form, i.e. $\Psi_{n{\bf k}}({\bf r})=[\psi^{\uparrow}_{n{\bf k}}({\bf r}),\psi^{\downarrow}_{n{\bf k}}({\bf r})]$. In this work, we have directly calculated the exchange self-energy matrix elements in real space. We have found that within the framework of the full-potential LAPW method, such elements cannot be evaluated efficiently in reciprocal space due to the extremely slow convergence with respect to the reciprocal lattice vectors. It is important to point out that when computing exchange self-energy elements, contributions from both the core and valence electrons have to be included to account for the important core-valence exchange process\cite{Ku_GW}.

Unlike $\Sigma^x$, the correlation self-energy within the KS basis can be evaluated efficiently in reciprocal space, and it reads

\begin{eqnarray}
 \Sigma^c_{n{\bf k}}(\omega)&=&\langle\Psi_{n{\bf k}}|\frac{i}{2\pi}\int d{\omega'}G(\omega+\omega')(W(\omega')-v)|\Psi_{n{\bf k}}\rangle \nonumber\\
&=&\frac{i}{2\pi}\sum\limits_{m}\int\limits_{BZ}\frac{d{\bf q}}{(2\pi)^3}\sum\limits_{{\bf G},{\bf G}'}v({\bf q}+{\bf G})[M^{{\bf k}-{\bf q}}_{mn}({\bf G},{\bf q})]^{*}\nonumber\\
&&\times  M^{{\bf k}-{\bf q}}_{mn}({\bf G'},{\bf q})\times\int d\omega'G^0_{m{\bf k}-{\bf q}}(\omega+\omega')\nonumber\\
&&\times \overline{\varepsilon}^{-1}_{{\bf GG'}}({\bf q},\omega'),\label{sigmac}
\end{eqnarray}
where $M^{{\bf k}}_{ij}({\bf G},{\bf q})=\langle\Psi_{i{\bf k}}|e^{-i({\bf q}+{\bf G})\cdot{\bf r}}|\Psi_{j{\bf k}+{\bf q}}\rangle$. This matrix contains all the information regarding the KS eigenfunctions. $v({\bf q}+{\bf G})=4\pi/|{\bf q}+{\bf G}|^2$ is the Fourier transform of the bare Coulomb potential. $G^{0}_{n{\bf k}}(\omega)=\frac{f_{n{\bf k}}}{\omega-\epsilon_{n{\bf k}}-i\eta}+\frac{1-f_{n{\bf k}}}{\omega-\epsilon_{n{\bf k}}+i\eta}$ is the non-interacting Green function written in ${\bf k}$ and $\omega$ space with $f_{n{\bf k}}$ being the occupation number of state $\Psi_{n{\bf k}}$. The dummy variable $m$ runs over all the valence and conduction bands, and the integration over ${\bf q}$ is within the first Brillouin Zone (BZ) of reciprocal space. $\overline{\varepsilon}^{-1}$ is the frequency-dependent term of the RPA inverse dielectric matrix.  Its element reads $\overline{\varepsilon}^{-1}_{{\bf GG'}}({\bf q},\omega)\equiv\varepsilon^{-1}_{{\bf GG'}}({\bf q},\omega)-\delta_{{\bf GG'}}$, where $\varepsilon_{{\bf GG'}}$ is the RPA dielectric matrix element, which reads
\begin{eqnarray}
\varepsilon_{{\bf GG'}}({\bf q},\omega)=\delta_{{\bf GG'}}-v({\bf q}+{\bf G})\chi^{KS}_{{\bf GG'}}({\bf q},\omega),
\end{eqnarray}
where $\chi^{KS}$ is the non-interacting density-density response function constructed using the KS eigenvalues and eigenfunctions,  
\begin{eqnarray}
\chi^{KS}_{{\bf GG'}}({\bf q},\omega)&=&\frac{1}{N_k\Omega}\sum\limits_{{\bf k}\in BZ}\sum\limits_{jj'}\frac{f_{j{\bf k}}-f_{j'{\bf k}+{\bf q}}}{\omega+\epsilon_{j{\bf k}}-\epsilon_{j'{\bf k}+{\bf q}}+i\eta}\nonumber\\
&&\times M^{k}_{jj'}({\bf q},{\bf G'})[M^{k}_{jj'}({\bf q},{\bf G})]^*.
\end{eqnarray}

Note that a convolution along the $\omega$ axis is needed in Eq.(\ref{sigmac}), which requires an inversion of the dielectric matrix $\varepsilon$ at all $\omega$ points. In this work, a direct evaluation of this convolution has been referred to as the NI method. In the current implementation of this method, $\varepsilon^{-1}$ is first computed on a set of $\omega$ points and linear interpolation is employed between any two neighbouring points for computing the convolution. 

Apart from the NI method, the PPA has also been implemented and tested. Within the PPA, the frequency dependence of $\varepsilon^{-1}$ is simplified and is approximated by a single-pole function of $\omega$ that reads
\begin{eqnarray}
\varepsilon^{-1}_{{\bf GG'}}({\bf q},\omega)&=&\delta_{{\bf GG'}}+\frac{R_{{\bf GG'}}({\bf q})}{\omega-\Omega_{{\bf GG'}}({\bf q})+i\delta}\nonumber\\
&&-\frac{R_{{\bf GG'}}({\bf q})}{\omega+\Omega_{{\bf GG'}}({\bf q})-i\delta},
\end{eqnarray}
where $R_{{\bf GG'}}({\bf q})$ and $\Omega_{{\bf GG'}}({\bf q})$ are the two matrices that need to be determined. We have adopted the fitting approach proposed by Godby and Needs (GN)\cite{GNPPM}. To do this, $\varepsilon^{-1}_{{\bf GG'}}({\bf q},\omega)$ is evaluated both at $\omega=0$ and $\omega=i\omega_p$ where $\hbar\omega_p$ is set as 1 Ha throughout this work.

Within the PPA, the correlation self-energy becomes
\begin{eqnarray}
\Sigma^c_{nk}(\omega)&=&\sum\limits_m\int\limits_{BZ}\frac{d{\bf q}}{(2\pi)^3}\sum\limits_{{\bf G},{\bf G'}}v({\bf q},{\bf G})[M^{{\bf k}-{\bf q}}_{mn}({\bf G},{\bf q})]^*\nonumber\\
&&\times M^{{\bf k}-{\bf q}}_{mn}({\bf G'},{\bf q})R_{{\bf GG'}}({\bf q})\times H^{m}_{{\bf k}-{\bf q}}(\omega),\\
H^{m}_{{\bf k}-{\bf q}}(\omega)&=&\frac{f_{m{\bf k}-{\bf q}}}{\omega-\epsilon_{m{\bf k}-{\bf q}}+\Omega_{{\bf GG'}}({\bf q})-i\eta}\nonumber\\
&&+\frac{1-f_{m{\bf k}-{\bf q}}}{\omega-\epsilon_{m{\bf k}-{\bf q}}-\Omega_{{\bf GG'}}({\bf q})+i\eta}.
\end{eqnarray}
 It is worth pointing out that when computing the correlation self-energy elements using either the NI method or the PPA, deep core electrons are neglected due to their small contribution.  In contrast, the relatively shallow core electrons are included and treated on the same footing as valence electrons.\cite{PhysRevB.66.125101}

\section{Model systems and computational details}
The group 14 nitride compounds in the cubic spinel phase, $\gamma$-M$_3$N$_4$, and in the $\beta$ phase, $\beta$-M$_3$N$_4$ have been studied in this work. Here, M= C, Si, Ge and Sn for $\gamma$-M$_3$N$_4$; M= Si, Ge and Sn for $\beta$-M$_3$N$_4$. The space groups for $\gamma$-M$_3$N$_4$ and $\beta$-M$_3$N$_4$ are $Fd$\,-3\,$m$ (227) and $P\,6_{3}/\,m$ (176), respectively. The $\gamma$-M$_3$N$_4$ has a face-centered cubic (FCC) structure while the $\beta$-M$_3$N$_4$ has a hexagonal structure. In both phases, there are six M and eight N atoms in the primitive cell (see Fig.\,\ref{cell}). 

\begin{table*}[t]
\caption{The lattice parameter $a$, direct band gap $E_g(\Gamma-\Gamma)$ values based on the DFT-LDA method, G$_0$W$_0$ with the PPA and G$_0$W$_0$ with the NI for $\gamma$-M$_3$N$_4$ compounds. Values in parenthesis are the experimental lattice parameters from Ref.\,\onlinecite{Zerrsi3n4}, \onlinecite{r-ge33} and \onlinecite{r-sn3}. The experimental band gaps are from Ref.\,\onlinecite{PhysRevLett.111.097402} and \onlinecite{PhysRevB.81.155207}.}
\begin{tabular}{ c c l c c c c}
\hline \hline
\noalign{\medskip}
$\quad$ System & $\quad$ $a\,(\aa\,)$ & & $\quad$ & $\quad$ $E_g(\Gamma-\Gamma$)\,(\eV\,) & $\quad$ & $\quad$   \\
\noalign{\medskip}
$\quad$  & $\quad$ & & $\quad$ LDA & $\quad$ G$_0$W$_0\,$(PPA) & $\quad$ G$_0$W$_0\,$(NI) & $\quad$ Expt.  \\
\noalign{\medskip}
\hline   
\noalign{\medskip}
 \ $\quad$ $\gamma$-C$_3$N$_4$ & $\quad$ 6.675 & & $\quad$ 1.17 & $\quad$ 1.96& $\quad$ 1.96 &  \\
 \ $\quad$ $\gamma$-Si$_3$N$_4$ & $\quad$ 7.697 & (7.80) & $\quad$ 3.35 & $\quad$ 4.87 & $\quad$ 4.89  & $\quad$ 4.6-5.0 \\ 
 \ $\quad$ $\gamma$-Ge$_3$N$_4$ & $\quad$ 8.211 & (8.2063) & $\quad$ 2.07 & $\quad$ 3.26 & $\quad$ 3.27 &  $\quad$ 3.3-3.7, 3.50 \\
 \ $\quad$ $\gamma$-Sn$_3$N$_4$ & $\quad$ 9.001 & (9.037) & $\quad$ 0.58 & $\quad$ 1.53 & $\quad$ 1.42 & $\quad$ 1.4-1.8 \\ 
 \noalign{\medskip}
 \hline \hline
\end{tabular}
\label{tab1}
\end{table*}

All the calculations have been performed using the modified version of the Elk full-potential LAPW code, in which we have implemented the G$_{0}$W$_{0}$ approach as described in Section II B. For the DFT calculations, the exchange-correlation functionals have been treated within the local density approximation (LDA). When expanding the intersitial potential, the maximum length of the reciprocal lattice vector $|{\bf G}|$ has been selected as 12 a.u. The first BZ has been sampled by a $4\times 4\times 4$ ${\bf k}$ mesh for the spinel-type structures and a $4\times 4\times 8$ ${\bf k}$ mesh for the $\beta$-phase structures. All the above parameters have carefully been tested for total energy convergence. The atomic structures for all the systems have been optimized until the force components on each atom are less than 0.05\,\eV/\aa. In the calculations concerning the Ge and Sn, the Ge $(3d)^{10}(4s)^2(4p)^2$ and Sn $(4p)^6 (4d)^{10}(5s)^2(5p)^2$ electrons have been treated as valence electrons.

For the GW calculations, 400 conduction bands have been used for computing the dielectric matrix and the Green function. Note that the GW band gap in all the systems only varies less than 0.04 eV when the number of conduction bands is reduced by a factor of two. This fast convergence regarding the band gap is likely due to the error cancellation between the quasiparticle energy corrections of the valence and conduction bands. We have used 61 frequency points for the NI method.

\section{Results}
\begin{figure}
\includegraphics[width=7.5cm]{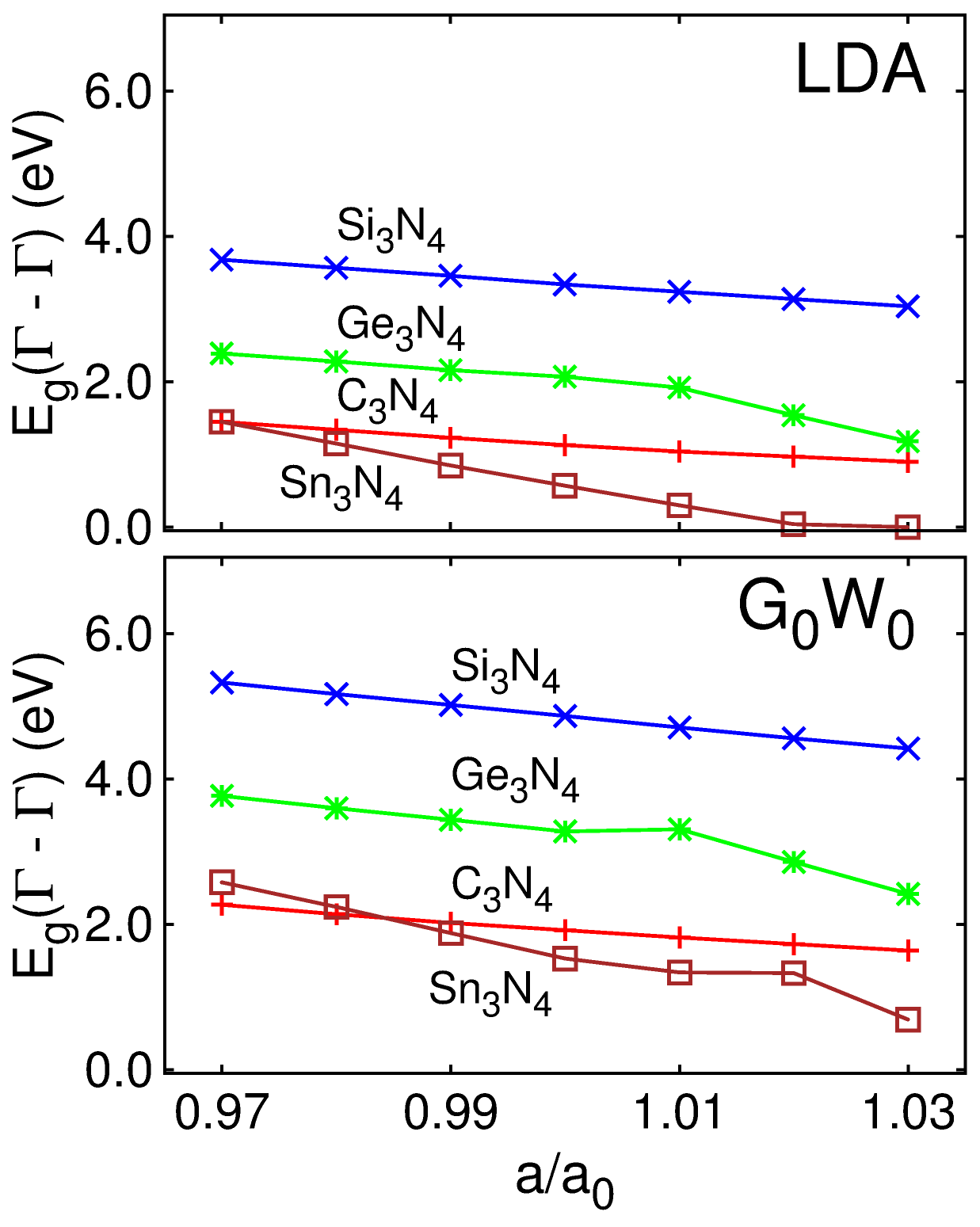}
\caption{(color online) The DFT-LDA and G$_0$W$_0$ band gaps as functions of the lattice parameter $a$ for $\gamma$-M$_3$N$_4$. $a_0$ is the optimized lattice parameter given in Table I.}\label{r_a_eg}
\end{figure}

\subsection{$\gamma$-M$_3$N$_4$}
First, we have investigated the $\gamma$-M$_3$N$_4$ compounds. The structural optimization has been carried out for each system and the resulting lattice parameters are given in Table I.  For $\gamma$-Si$_3$N$_4$, $\gamma$-Ge$_3$N$_4$ and $\gamma$-Sn$_3$N$_4$, they are in very good agreement with the experimental lattice parameters. The percentage difference is about 1$\%$ or less. 

\begin{figure*}
\includegraphics[width=14cm]{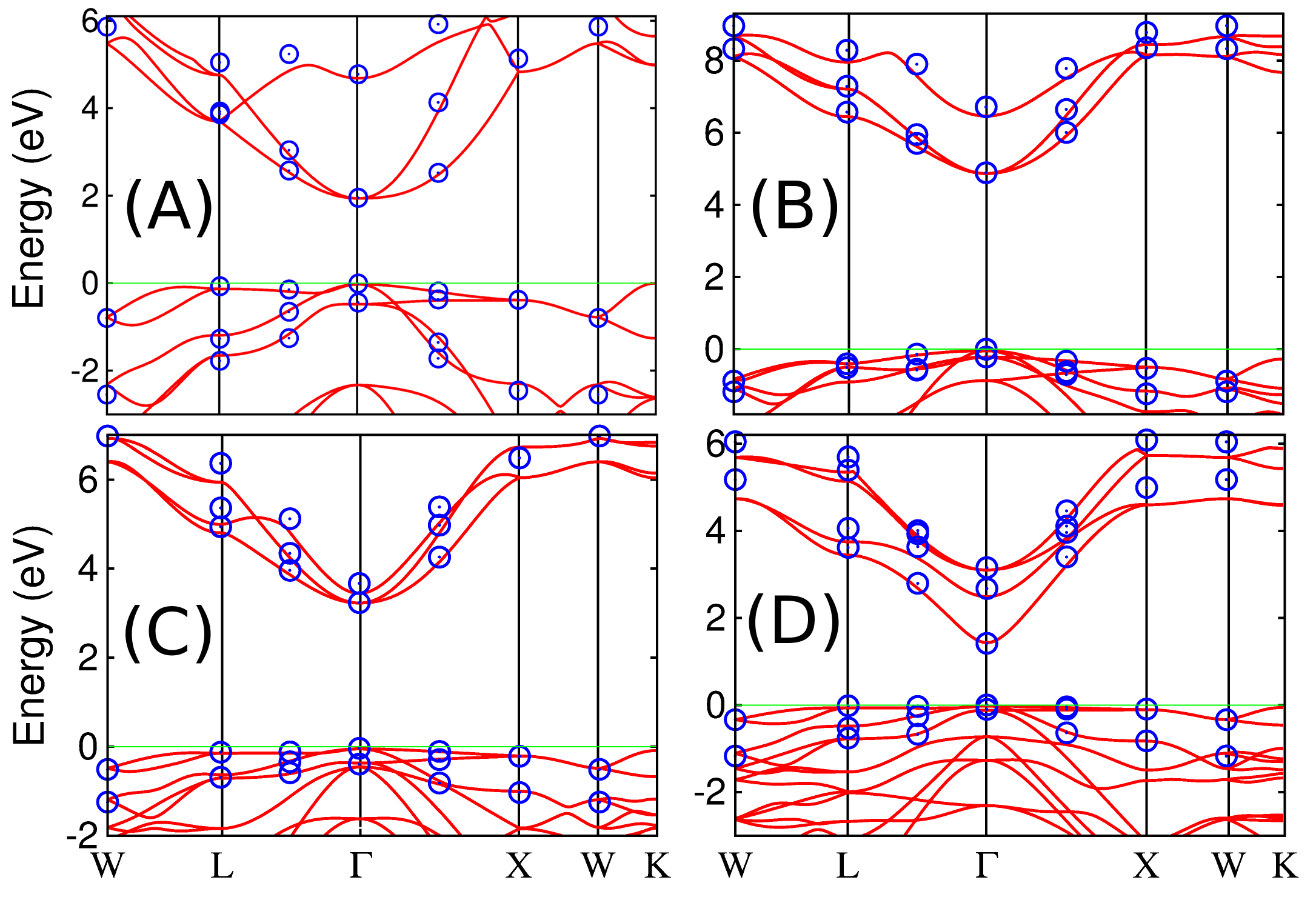}
\caption{(color online) Band gap corrected LDA band structures (red lines) for (A) $\gamma$-C$_3$N$_4$, (B) $\gamma$-Si$_3$N$_4$, (C) $\gamma$-Ge$_3$N$_4$ and (D) $\gamma$-Sn$_3$N$_4$. The conduction bands are shifted upward using the G$_0$W$_0$ gap values. The corresponding quasiparticle energy corrections (4 valence and 4 conduction bands near the band gap) with the NI are illustrated as blue circles. The VBM is indicated by the green line.}\label{band_rl}
\end{figure*}

The DFT-LDA electronic band gaps have been computed and the direct gap values at $\Gamma$, $E_g(\Gamma-\Gamma)$ are illustrated in Table I. For $\gamma$-Si$_3$N$_4$, $\gamma$-Ge$_3$N$_4$ and $\gamma$-Sn$_3$N$_4$, we have found that the minimal band gap is always direct at $\Gamma$. For the hypothetical $\gamma$-C$_3$N$_4$, however, the band gap is found indirect and is 0.17\,\eV\,smaller than $E_g(\Gamma-\Gamma)$. In this case, the conduction band minimum (CBM) is located at $\Gamma$ whereas the valence band maximum (VBM) is near the $k$ point (0.75, 0.25, 0.00), expressed in fractional coordinates of the primitive reciprocal lattice vectors. This is contrary to other theoretical findings\cite{PhysRevB.63.064102} that the minimal band gap in this system is found direct at $\Gamma$. Moreover,  our DFT-LDA $E_g(\Gamma-\Gamma)$ for $\gamma$-C$_3$N$_4$, $\gamma$-Si$_3$N$_4$ and $\gamma$-Ge$_3$N$_4$ are very similar to those obtained by the plane-wave based pseudopotential methods, with the difference always less than 0.2\,\eV. This indicates that pseudopotential methods yield only small errors in the  calculations for these compounds. For $\gamma$-Sn$_3$N$_4$, our band gap is about 0.7\,\eV\,less than that from other theoretical  calculations\cite{PhysRevB.63.064102}.

The DFT-LDA band gaps are consistently underestimated, and are at least 0.8\eV\,smaller compared with the available experimental counterparts. To improve accuracy, we have performed the single-shot G$_0$W$_0$ calculations using the PPA or the NI method for the band gaps (see Table I). For $\gamma$-C$_3$N$_4$, $\gamma$-Si$_3$N$_4$ and $\gamma$-Ge$_3$N$_4$, the GW band gaps using the PPA are remarkably close to those using the NI method. For the $\gamma$-Sn$_3$N$_4$, the band gap difference is 0.11\,\eV\,between the two methods. Overall, our results agree with the previous studies which claim that the GN-PPA results can consistently match the NI counterparts\cite{PhysRevB.84.241201, PhysRevB.88.125205}. Using either method, the computed band gaps are in good agreement with experiments (see Table I).

\begin{table*}[t]
\caption{The lattice parameters $a$ and $c$, calculated band gaps $E_g$ based on the DFT-LDA method, G$_0$W$_0$ with the PPA and G$_0$W$_0$ with the NI for $\beta$-M$_3$N$_4$ compounds. Values in parenthesis are the experimental lattice parameters from Ref.\,\onlinecite{bsi} and \onlinecite{bge}. The experimental band gap is from Ref.\,\onlinecite{bge2}.}
\begin{tabular}{ c c l c l c c c c}
\hline \hline
\noalign{\medskip}
$\quad$ System & $\quad$ $a\,(\aa\,)$ & & $\quad$ $c\,(\aa\,)$ & & $\quad$ & $\quad$ $E_g$\,(\eV\,) & $\quad$ & $\quad$  \\
\noalign{\medskip}
$\quad$ & $\quad$ & & $\quad$ & & $\quad$ LDA & $\quad$ G$_0$W$_0\,$(PPA) & $\quad$ G$_0$W$_0\,$(NI) & $\quad$ Expt. \\
\noalign{\medskip}
\hline   
\noalign{\medskip}
 $\quad$ $\beta$-Si$_3$N$_4$ & $\quad$ 7.576 & (7.607) & $\quad$ 2.892 & (2.911) & $\quad$ 4.19 & $\quad$ 6.06 & $\quad$ 6.06 &  \\ 
 $\quad$ $\beta$-Ge$_3$N$_4$ & $\quad$ 8.063 & (8.038) & $\quad$ 3.084 & (3.074) & $\quad$ 2.03 & $\quad$ 3.60 & $\quad$ 3.59 & 4.4-4.8 \\
 $\quad$ $\beta$-Sn$_3$N$_4$ & $\quad$ 8.814 &  & $\quad$ 3.418 &  & $\quad$ 0.13 & $\quad$ 1.04 & $\quad$ 0.98 & \\
 \noalign{\medskip}
 \hline \hline
\end{tabular}
\label{tab2}
\end{table*}

To understand the impact of the lattice parameter $a$ on the electronic band gap, we have performed calculations at both the DFT-LDA and the G$_0$W$_0$ levels. In these calculations, $a$ varies between 0.97 to 1.03 times the optimized lattice parameter, $a_0$ given in Table I. The results are illustrated in Fig.\,\ref{r_a_eg}. At each given $a$, the DFT-LDA structural optimization has been first employed for the lowest total energy. The same lattice parameter has then been adopted for the subsequent band gap calculations. Our results suggest that the band gap at $\Gamma$ $E_g(\Gamma-\Gamma)$ is very sensitive to the lattice parameter. For $\gamma$-C$_3$N$_4$ and $\gamma$-Si$_3$N$_4$, the DFT-LDA results show that $E_g(\Gamma-\Gamma)$ linearly decreases along with the increase of $a$. For $\gamma$-C$_3$N$_4$, the trend for the indirect band gap is very similar to the one demonstrated and is not shown here. Within the range of $a/a_0$, $E_g(\Gamma-\Gamma)$ varies between 0.9 and 1.45 eV for $\gamma$-C$_3$N$_4$ and between 3.04 and 3.68 eV for $\gamma$-Si$_3$N$_4$. Upon the GW corrections, the curves for these two systems are shifted upward compared with the DFT-LDA results, which suggests that the band gap corrections due to the GW method stays almost the same as $a$ changes. For $\gamma$-Ge$_3$N$_4$, the DFT-LDA band gap also decreases linearly when $a$ increases. Within the range of $a$, the DFT-LDA band gap varies between 1.18 and 2.39 eV.  Note that the slope changes when $a/a_0$ becomes 1.01. Further investigation suggests that this variation stems from the  symmetry change at the CBM: the CBM is triply degenerate when $a/a_0$ is smaller than 1.01, and becomes singly degenerate otherwise. Upon the GW correction, the curve has a similar trend as its DFT-LDA counterpart, and a kink appears at $a/a_0=1.01$. The GW band gap correction varies little at different $a$ values. For $\gamma$-Sn$_3$N$_4$, the DFT-LDA band gap  decreases linearly as $a/a_0$ increases until $a/a_0$ equals 1.02, at which point the band gap is reduced to 0.04 eV. At $a/a_0=1.03$, the band gap becomes zero. The band gap variation in this range is 1.45 eV, which is 250\,\% more than the band gap at $a=a_0$. With the GW correction included, the band gap monotonically decreases when $a$ increases. Note that the GW correction results in a non-zero band gap when $a/a_0=1.03$, despite the zero gap predicted by the DFT-LDA calculations. Also, the band gap correction at $1.03\,a_0$ is smaller compared with those at other $a$ values. This is due to the sign change (from negative to positive) of the correlation self-energy $\Sigma_c$ at the CBM.

The DFT-LDA band structures for all the compounds at $a=a_0$ are demonstrated in Fig.\,\ref{band_rl}, in which the conduction bands have been shifted upward using the corresponding GW-NI band gap values. With the use of the NI method, the GW quasiparticle energies for the four valence bands and four conduction bands near the band gap have also been shown in the figure (blue circles). We have found that after shifting the conduction bands using the GW-corrected band gap, the quasiparticle energies in the valence band are very close to their DFT counterparts, whereas in the conduction band the difference remains small only near the $\Gamma$ point. Since the VBM and CBM are both located at $\Gamma$ for the $\gamma$-Si$_3$N$_4$, $\gamma$-Ge$_3$N$_4$ and $\gamma$-Sn$_3$N$_4$, this indicates that the difference of the optical properties predicted by the DFT and by the GW method is mainly due to the band gap corrections. Our results also imply that the band gap property (direct or indirect) is conserved after the GW corrections have been applied.

\begin{figure}
\includegraphics[width=7.5cm]{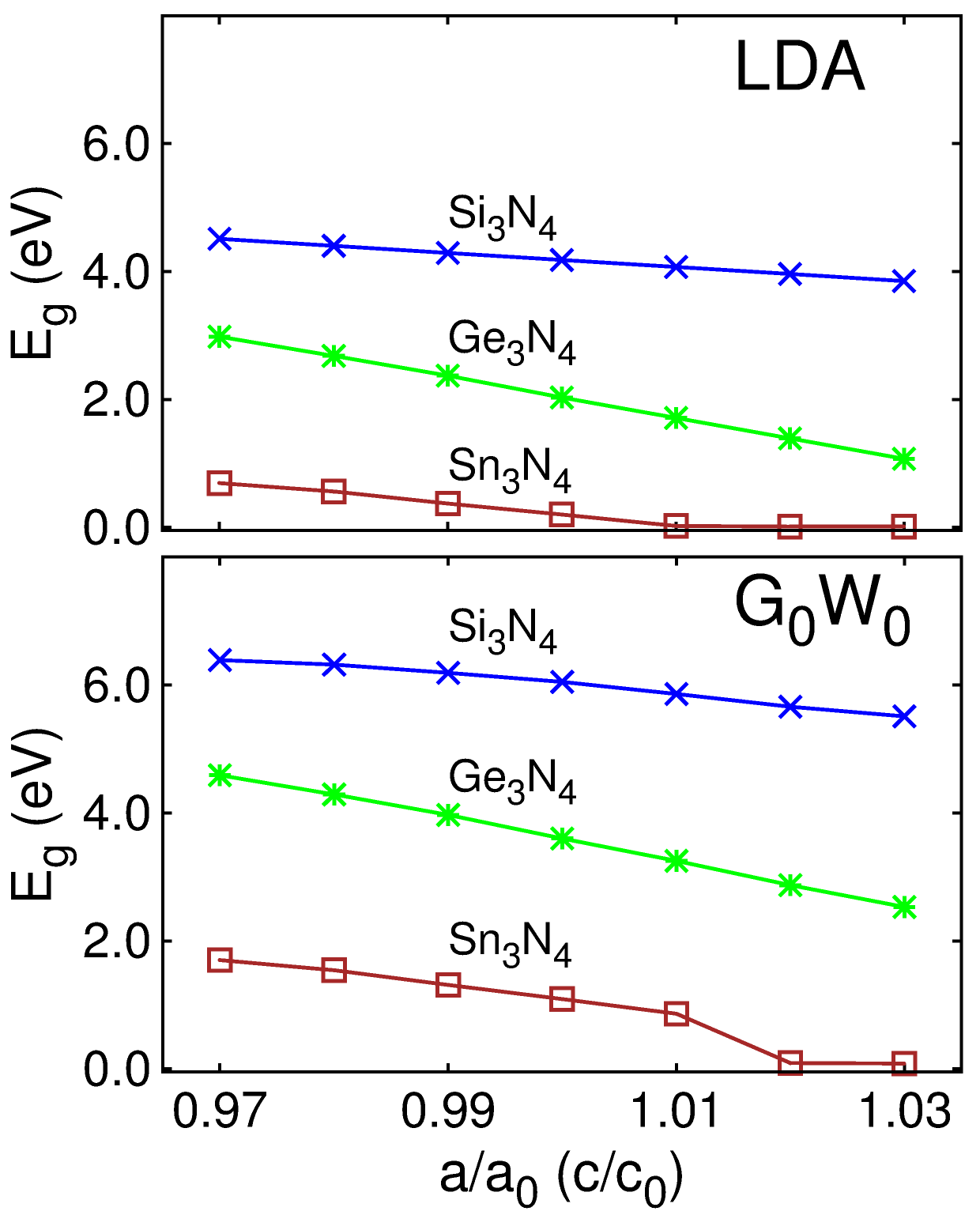}
\caption{(color online) The DFT-LDA and G$_0$W$_0$ band gaps as functions of the lattice parameters $a$ and $c$ for the $\beta$-M$_3$N$_4$. $a_0$ and $c_0$ are the optimized lattice parameters given in Table II.}\label{b_a_eg}
\end{figure}

\subsection{$\beta$-M$_3$N$_4$}
Next, we have studied the $\beta$-M$_3$N$_4$ compounds with M=Si, Ge and Sn. Based on the available experimental data, $\beta$-Sn$_3$N$_4$ remains hypothetical whereas the other two have been synthesized successfully.  As a first step, their atomic structures have been optimized and the resulting lattice parameters $a$ and $c$ are given in Table II.  For $\beta$-Si$_3$N$_4$ and $\beta$-Ge$_3$N$_4$, these parameters are in very good agreement with experiment (see Table II). The percentage difference is always less than 1\,\%.

\begin{figure}
\includegraphics[width=7.5cm]{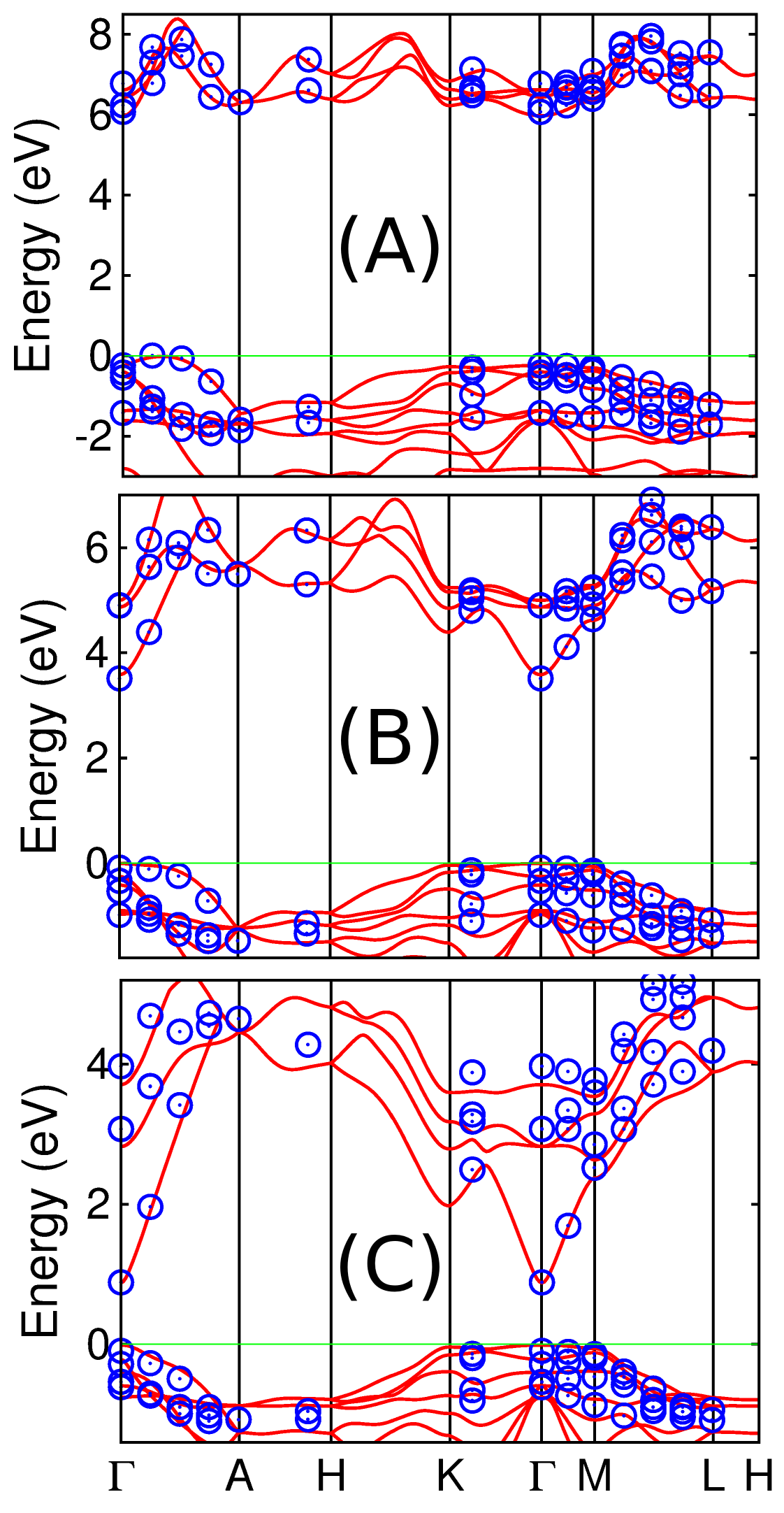}
\caption{(color online) Band gap corrected LDA band structures (red lines) for (A) $\beta$-Si$_3$N$_4$, (B) $\beta$-Ge$_3$N$_4$ and (C) $\beta$-Sn$_3$N$_4$. The conduction bands are shifted upward using the G$_0$W$_0$ gap values. The corresponding quasiparticle energy corrections (4 valence and 4 conduction bands near the band gap) with the NI are illustrated as blue circles. The VBM is indicated by the green line.}\label{band_b}
\end{figure}

The DFT-LDA band gaps are shown in Table II. For $\beta$-Si$_3$N$_4$, the band gap at $\Gamma$ is 4.39\,\eV. This system is known as an indirect band-gap semiconductor. Based on our calculations, the minimal band gap is estimated as 4.19\,\eV, in which the CBM and VBM are located at $\Gamma$ and the $k$-point (0, 0, 0.125) in fractional coordinates. For both the $\beta$-Ge$_3$N$_4$ and the hypothetical $\beta$-Sn$_3$N$_4$, the minimal band gap is at $\Gamma$. Based on our calculations, the DFT-LDA band gap for the $\beta$-Sn$_3$N$_4$  has been obtained as 0.13\,\eV. Upon the GW corrections using either the PPA or NI method, the band gaps are much larger than those DFT-LDA counterparts. For $\beta$-Ge$_3$N$_4$, the DFT-LDA band gap is 2.03\,\eV, which is much smaller than Gao \textit{et al.}'s calculated value of 2.92\,\eV\cite{Gao2013292}. This may be due to the different lattice parameters $a$ and $c$ used in the calculations: we have found that our $a$ and $c$ are both about 2\% bigger than those computed using the pseudopotential method. If using 2\% smaller all-electron lattice parameters for the DFT-LDA calculation, the band gap calculated by our method is 2.68 eV that matches well with the pseudopotential calculations. This suggests a strong dependence of the band gap on $a$ and $c$ (see Fig.\,\ref{b_a_eg}) as seen for the other nitride compounds. Note that the corresponding GW band gap is 3.60\,\eV, which is smaller than 4.293\,\eV\,given in Gao \textit{et al.}'s results\cite{Gao2013292}, and is also 0.8\,\eV\,smaller than the experimental value. Based on the fact that our all-electron lattice parameters are in very good agreement with experiments, the band gap difference between our results and the experiments may result from the exclusion of self-consistency in the G$_0$W$_0$ method, and the band gap corrections may depend on the starting point, e.g. the LDA KS eigenvalues and eigenfunctions. A self-consistent GW approach can eliminate this dependency and may reduce the gap between the calculated and experimental results. More studies on this issue may be addressed in the future.

Next, we have studied the dependence of band gap on the lattice parameters $a$ and $c$ for these compounds. In our calculations, $a$ ($c$) varies between 0.97 and 1.03 with respect to $a_0$ ($c_0$), the optimized lattice parameter given in Table II. The $c/a$ ratio has been kept unchanged in the calculations.  Again, the structural optimization has been carried out at each selected pair of $a$ and $c$. Then the band gap at both the DFT-LDA and G$_0$W$_0$ levels has been obtained. The results are illustrated in Fig.\,\ref{b_a_eg}. For $\beta$-Si$_3$N$_4$ and $\beta$-Ge$_3$N$_4$, the DFT-LDA band gap linearly decreases as $a$ ($c$) increases. The corresponding GW band gaps follow the same trend, with an overall shift upward compared with the DFT-LDA results. This again suggests that the GW correction to the DFT-LDA band gap is weakly dependent on the varying $a$ ($c$) within this range. For $\beta$-Sn$_3$N$_4$, the DFT-LDA band gap presents a decreasing trend when the lattice parameters increase before $a/a_0=1.01$, and the gap value approaches zero when $a/a_0$ greater than 1.01. Our calculations indicate that the location of the VBM is changed when $a/a_0\ge 1.01$, which leads to an indirect band gap for this system. The corresponding GW band gap at $a/a_0=1.02$ and 1.03 is less than 0.1\,\eV. 

The DFT-LDA band structures for the compounds at $a=a_0$ ($c=c_0$) have been illustrated in Fig. \ref{band_b}. Similarly, the conduction bands have been shifted upward using the GW-NI gap value. The GW quasiparticle energies for the four valence and four conduction bands are shown in the figure as well. In each case, the GW corrections are very small in the valence band, whereas the corrections in the conduction band are small near the CBM at $\Gamma$. Our GW calculations also suggest direct band gaps at $\Gamma$ for the $\beta$-Ge$_3$N$_4$ and $\beta$-Sn$_3$N$_4$.

\section{Conclusions}
To summarize, we have studied the electronic properties of the group 14 nitride compounds in the spinel phase  $\gamma$-M$_3$N$_4$ (M=C, Si, Ge and Sn) and the $\beta$ phase $\beta$-M$_3$N$_4$ (M=Si, Ge and Sn) within the framework of the full-potential LAPW method. The lattice parameters obtained by the structural optimization are in good agreement with experiments. For the spinel-type structures $\gamma$-M$_3$N$_4$, direct band gaps at $\Gamma$ have been found for M=Si, Ge and Sn, whereas the band gap is indirect for the hypothetical $\gamma$-C$_3$N$_4$. Using single-shot G$_0$W$_0$ corrections to account for  many-body interactions, the band gaps agree very well with the experimental data for the spinel structures. Moreover, our calculations at the DFT-LDA and G$_0$W$_0$ levels have shown a strong dependence of the band gap on the lattice parameters in both phases. In particular, the band gap always decreases as the lattice parameters increase. We have also compared our GW results based on the PPA to those obtained by the NI method. We have found that the GW-PPA band gaps for all the structures studied are consistently close to the GW-NI results, with a difference less than 0.11\,\eV. 

Our GW calculations include the minimal approximations except for those in the G$_0$W$_0$ approach. In particular, the pseudopotential method, which has been widely used for the first-principles studies, is not required here. Our theoretical results for the predicted $\gamma$-C$_3$N$_4$ and $\beta$-Sn$_3$N$_4$ can provide guidance for future experiments. The spinel-type nitrides $\gamma$-M$_3$N$_4$ (M=Si, Ge and Sn) in our study are novel high-pressure compounds that are very promising for optoelectronic applications. 

\section{Acknowledgements}
This work is supported by NSF/DMR-0804407 and DOE/BES-DE-FG02-02ER45995. We acknowledge Dr. Yun-Peng Wang for helpful discussions. The calculations have been performed at NERSC and UF-HPC Center.

\bibliography{nitride_paper_v4}

\begin{thebibliography}{45}
\expandafter\ifx\csname natexlab\endcsname\relax\def\natexlab#1{#1}\fi
\expandafter\ifx\csname bibnamefont\endcsname\relax
  \def\bibnamefont#1{#1}\fi
\expandafter\ifx\csname bibfnamefont\endcsname\relax
  \def\bibfnamefont#1{#1}\fi
\expandafter\ifx\csname citenamefont\endcsname\relax
  \def\citenamefont#1{#1}\fi
\expandafter\ifx\csname url\endcsname\relax
  \def\url#1{\texttt{#1}}\fi
\expandafter\ifx\csname urlprefix\endcsname\relax\def\urlprefix{URL }\fi
\providecommand{\bibinfo}[2]{#2}
\providecommand{\eprint}[2][]{\url{#2}}

\bibitem[{\citenamefont{Ching et~al.}(2000)\citenamefont{Ching, Ouyang, and
  Gale}}]{PhysRevB.61.8696}
\bibinfo{author}{\bibfnamefont{W.~Y.} \bibnamefont{Ching}},
  \bibinfo{author}{\bibfnamefont{L.}~\bibnamefont{Ouyang}}, \bibnamefont{and}
  \bibinfo{author}{\bibfnamefont{J.~D.} \bibnamefont{Gale}},
  \bibinfo{journal}{Phys. Rev. B} \textbf{\bibinfo{volume}{61}},
  \bibinfo{pages}{8696} (\bibinfo{year}{2000}).

\bibitem[{\citenamefont{Kruger et~al.}(1997)\citenamefont{Kruger, Nguyen, Li,
  Caldwell, Manghnani, and Jeanloz}}]{PhysRevB.55.3456}
\bibinfo{author}{\bibfnamefont{M.~B.} \bibnamefont{Kruger}},
  \bibinfo{author}{\bibfnamefont{J.~H.} \bibnamefont{Nguyen}},
  \bibinfo{author}{\bibfnamefont{Y.~M.} \bibnamefont{Li}},
  \bibinfo{author}{\bibfnamefont{W.~A.} \bibnamefont{Caldwell}},
  \bibinfo{author}{\bibfnamefont{M.~H.} \bibnamefont{Manghnani}},
  \bibnamefont{and} \bibinfo{author}{\bibfnamefont{R.}~\bibnamefont{Jeanloz}},
  \bibinfo{journal}{Phys. Rev. B} \textbf{\bibinfo{volume}{55}},
  \bibinfo{pages}{3456} (\bibinfo{year}{1997}).

\bibitem[{\citenamefont{Bradley et~al.}(1966)\citenamefont{Bradley, Munro, and
  Whitfield}}]{Bradley19661803}
\bibinfo{author}{\bibfnamefont{R.}~\bibnamefont{Bradley}},
  \bibinfo{author}{\bibfnamefont{D.}~\bibnamefont{Munro}}, \bibnamefont{and}
  \bibinfo{author}{\bibfnamefont{M.}~\bibnamefont{Whitfield}},
  \bibinfo{journal}{J. Inorg. Nucl. Chem.} \textbf{\bibinfo{volume}{28}},
  \bibinfo{pages}{1803 } (\bibinfo{year}{1966}).

\bibitem[{\citenamefont{Pham et~al.}(2011)\citenamefont{Pham, Li, Shankar,
  Gygi, and Galli}}]{PhysRevB.84.045308}
\bibinfo{author}{\bibfnamefont{T.~A.} \bibnamefont{Pham}},
  \bibinfo{author}{\bibfnamefont{T.}~\bibnamefont{Li}},
  \bibinfo{author}{\bibfnamefont{S.}~\bibnamefont{Shankar}},
  \bibinfo{author}{\bibfnamefont{F.}~\bibnamefont{Gygi}}, \bibnamefont{and}
  \bibinfo{author}{\bibfnamefont{G.}~\bibnamefont{Galli}},
  \bibinfo{journal}{Phys. Rev. B} \textbf{\bibinfo{volume}{84}},
  \bibinfo{pages}{045308} (\bibinfo{year}{2011}).

\bibitem[{\citenamefont{Liu and Cohen}(1990)}]{PhysRevB.41.10727}
\bibinfo{author}{\bibfnamefont{A.~Y.} \bibnamefont{Liu}} \bibnamefont{and}
  \bibinfo{author}{\bibfnamefont{M.~L.} \bibnamefont{Cohen}},
  \bibinfo{journal}{Phys. Rev. B} \textbf{\bibinfo{volume}{41}},
  \bibinfo{pages}{10727} (\bibinfo{year}{1990}).

\bibitem[{\citenamefont{Borgen and Seip}(1961)}]{bsi}
\bibinfo{author}{\bibfnamefont{O.}~\bibnamefont{Borgen}} \bibnamefont{and}
  \bibinfo{author}{\bibfnamefont{H.~M.} \bibnamefont{Seip}},
  \bibinfo{journal}{Acta Chem. Scand.} \textbf{\bibinfo{volume}{15}},
  \bibinfo{pages}{1789} (\bibinfo{year}{1961}).

\bibitem[{\citenamefont{Ruddlesden and Popper}(1958)}]{bge}
\bibinfo{author}{\bibfnamefont{S.~N.} \bibnamefont{Ruddlesden}}
  \bibnamefont{and} \bibinfo{author}{\bibfnamefont{P.}~\bibnamefont{Popper}},
  \bibinfo{journal}{Acta Cryst.} \textbf{\bibinfo{volume}{11}},
  \bibinfo{pages}{465} (\bibinfo{year}{1958}).

\bibitem[{\citenamefont{Zerr et~al.}(1999)\citenamefont{Zerr, Miehe, Serghiou,
  Schwarz, Kroke, Riedel, Fuesz, Kroll, and Boehler}}]{Zerrsi3n4}
\bibinfo{author}{\bibfnamefont{A.}~\bibnamefont{Zerr}},
  \bibinfo{author}{\bibfnamefont{G.}~\bibnamefont{Miehe}},
  \bibinfo{author}{\bibfnamefont{G.}~\bibnamefont{Serghiou}},
  \bibinfo{author}{\bibfnamefont{M.}~\bibnamefont{Schwarz}},
  \bibinfo{author}{\bibfnamefont{E.}~\bibnamefont{Kroke}},
  \bibinfo{author}{\bibfnamefont{R.}~\bibnamefont{Riedel}},
  \bibinfo{author}{\bibfnamefont{H.}~\bibnamefont{Fuesz}},
  \bibinfo{author}{\bibfnamefont{P.}~\bibnamefont{Kroll}}, \bibnamefont{and}
  \bibinfo{author}{\bibfnamefont{R.}~\bibnamefont{Boehler}},
  \bibinfo{journal}{Nature} \textbf{\bibinfo{volume}{400}},
  \bibinfo{pages}{340} (\bibinfo{year}{1999}).

\bibitem[{\citenamefont{Yang et~al.}(2007)\citenamefont{Yang, Wang, Feng, Peng,
  and Sun}}]{Yang_ge3n4}
\bibinfo{author}{\bibfnamefont{M.}~\bibnamefont{Yang}},
  \bibinfo{author}{\bibfnamefont{S.~J.} \bibnamefont{Wang}},
  \bibinfo{author}{\bibfnamefont{Y.~P.} \bibnamefont{Feng}},
  \bibinfo{author}{\bibfnamefont{G.~W.} \bibnamefont{Peng}}, \bibnamefont{and}
  \bibinfo{author}{\bibfnamefont{Y.~Y.} \bibnamefont{Sun}},
  \bibinfo{journal}{J. Appl. Phys.} \textbf{\bibinfo{volume}{102}}
  (\bibinfo{year}{2007}).

\bibitem[{\citenamefont{Dong et~al.}(2000)\citenamefont{Dong, Sankey, Deb,
  Wolf, and McMillan}}]{PhysRevB.61.11979}
\bibinfo{author}{\bibfnamefont{J.}~\bibnamefont{Dong}},
  \bibinfo{author}{\bibfnamefont{O.~F.} \bibnamefont{Sankey}},
  \bibinfo{author}{\bibfnamefont{S.~K.} \bibnamefont{Deb}},
  \bibinfo{author}{\bibfnamefont{G.}~\bibnamefont{Wolf}}, \bibnamefont{and}
  \bibinfo{author}{\bibfnamefont{P.~F.} \bibnamefont{McMillan}},
  \bibinfo{journal}{Phys. Rev. B} \textbf{\bibinfo{volume}{61}},
  \bibinfo{pages}{11979} (\bibinfo{year}{2000}).

\bibitem[{\citenamefont{Ching et~al.}(2002)\citenamefont{Ching, Mo, Ouyang,
  Rulis, Tanaka, and Yoshiya}}]{JACE75}
\bibinfo{author}{\bibfnamefont{W.-Y.} \bibnamefont{Ching}},
  \bibinfo{author}{\bibfnamefont{S.-D.} \bibnamefont{Mo}},
  \bibinfo{author}{\bibfnamefont{L.}~\bibnamefont{Ouyang}},
  \bibinfo{author}{\bibfnamefont{P.}~\bibnamefont{Rulis}},
  \bibinfo{author}{\bibfnamefont{I.}~\bibnamefont{Tanaka}}, \bibnamefont{and}
  \bibinfo{author}{\bibfnamefont{M.}~\bibnamefont{Yoshiya}},
  \bibinfo{journal}{J. Am. Chem. Soc.} \textbf{\bibinfo{volume}{85}},
  \bibinfo{pages}{75} (\bibinfo{year}{2002}).

\bibitem[{\citenamefont{Ching and Rulis}(2006)}]{PhysRevB.73.045202}
\bibinfo{author}{\bibfnamefont{W.~Y.} \bibnamefont{Ching}} \bibnamefont{and}
  \bibinfo{author}{\bibfnamefont{P.}~\bibnamefont{Rulis}},
  \bibinfo{journal}{Phys. Rev. B} \textbf{\bibinfo{volume}{73}},
  \bibinfo{pages}{045202} (\bibinfo{year}{2006}).

\bibitem[{\citenamefont{Mo et~al.}(1999)\citenamefont{Mo, Ouyang, Ching,
  Tanaka, Koyama, and Riedel}}]{PhysRevLett.83.5046}
\bibinfo{author}{\bibfnamefont{S.-D.} \bibnamefont{Mo}},
  \bibinfo{author}{\bibfnamefont{L.}~\bibnamefont{Ouyang}},
  \bibinfo{author}{\bibfnamefont{W.~Y.} \bibnamefont{Ching}},
  \bibinfo{author}{\bibfnamefont{I.}~\bibnamefont{Tanaka}},
  \bibinfo{author}{\bibfnamefont{Y.}~\bibnamefont{Koyama}}, \bibnamefont{and}
  \bibinfo{author}{\bibfnamefont{R.}~\bibnamefont{Riedel}},
  \bibinfo{journal}{Phys. Rev. Lett.} \textbf{\bibinfo{volume}{83}},
  \bibinfo{pages}{5046} (\bibinfo{year}{1999}).

\bibitem[{\citenamefont{Kuwabara et~al.}(2008)\citenamefont{Kuwabara,
  Matsunaga, and Tanaka}}]{PhysRevB.78.064104}
\bibinfo{author}{\bibfnamefont{A.}~\bibnamefont{Kuwabara}},
  \bibinfo{author}{\bibfnamefont{K.}~\bibnamefont{Matsunaga}},
  \bibnamefont{and} \bibinfo{author}{\bibfnamefont{I.}~\bibnamefont{Tanaka}},
  \bibinfo{journal}{Phys. Rev. B} \textbf{\bibinfo{volume}{78}},
  \bibinfo{pages}{064104} (\bibinfo{year}{2008}).

\bibitem[{\citenamefont{Boyko et~al.}(2013)\citenamefont{Boyko, Hunt, Zerr, and
  Moewes}}]{PhysRevLett.111.097402}
\bibinfo{author}{\bibfnamefont{T.~D.} \bibnamefont{Boyko}},
  \bibinfo{author}{\bibfnamefont{A.}~\bibnamefont{Hunt}},
  \bibinfo{author}{\bibfnamefont{A.}~\bibnamefont{Zerr}}, \bibnamefont{and}
  \bibinfo{author}{\bibfnamefont{A.}~\bibnamefont{Moewes}},
  \bibinfo{journal}{Phys. Rev. Lett.} \textbf{\bibinfo{volume}{111}},
  \bibinfo{pages}{097402} (\bibinfo{year}{2013}).

\bibitem[{\citenamefont{Pradhan et~al.}(2010)\citenamefont{Pradhan, Kumar, Deb,
  Waghmare, and Narayana}}]{PhysRevB.82.144112}
\bibinfo{author}{\bibfnamefont{G.~K.} \bibnamefont{Pradhan}},
  \bibinfo{author}{\bibfnamefont{A.}~\bibnamefont{Kumar}},
  \bibinfo{author}{\bibfnamefont{S.~K.} \bibnamefont{Deb}},
  \bibinfo{author}{\bibfnamefont{U.~V.} \bibnamefont{Waghmare}},
  \bibnamefont{and} \bibinfo{author}{\bibfnamefont{C.}~\bibnamefont{Narayana}},
  \bibinfo{journal}{Phys. Rev. B} \textbf{\bibinfo{volume}{82}},
  \bibinfo{pages}{144112} (\bibinfo{year}{2010}).

\bibitem[{\citenamefont{Huang and Feng}(2004)}]{Huang_sn3n4}
\bibinfo{author}{\bibfnamefont{M.}~\bibnamefont{Huang}} \bibnamefont{and}
  \bibinfo{author}{\bibfnamefont{Y.~P.} \bibnamefont{Feng}},
  \bibinfo{journal}{J. Appl. Phys.} \textbf{\bibinfo{volume}{96}}
  (\bibinfo{year}{2004}).

\bibitem[{\citenamefont{Scotti et~al.}(1999)\citenamefont{Scotti, Kockelmann,
  Senker, Traßel, and Jacobs}}]{r-sn3}
\bibinfo{author}{\bibfnamefont{N.}~\bibnamefont{Scotti}},
  \bibinfo{author}{\bibfnamefont{W.}~\bibnamefont{Kockelmann}},
  \bibinfo{author}{\bibfnamefont{J.}~\bibnamefont{Senker}},
  \bibinfo{author}{\bibfnamefont{S.}~\bibnamefont{Traßel}}, \bibnamefont{and}
  \bibinfo{author}{\bibfnamefont{H.}~\bibnamefont{Jacobs}},
  \bibinfo{journal}{Z. Anorg. Allg. Chem.} \textbf{\bibinfo{volume}{625}},
  \bibinfo{pages}{1435} (\bibinfo{year}{1999}).

\bibitem[{\citenamefont{Serghiou
  et~al.}(1999{\natexlab{a}})\citenamefont{Serghiou, Miehe, Tschauner, Zerr,
  and Boehler}}]{r-ge3}
\bibinfo{author}{\bibfnamefont{G.}~\bibnamefont{Serghiou}},
  \bibinfo{author}{\bibfnamefont{G.}~\bibnamefont{Miehe}},
  \bibinfo{author}{\bibfnamefont{O.}~\bibnamefont{Tschauner}},
  \bibinfo{author}{\bibfnamefont{A.}~\bibnamefont{Zerr}}, \bibnamefont{and}
  \bibinfo{author}{\bibfnamefont{R.}~\bibnamefont{Boehler}},
  \bibinfo{journal}{J. Chem. Phys.} \textbf{\bibinfo{volume}{111}}
  (\bibinfo{year}{1999}{\natexlab{a}}).

\bibitem[{\citenamefont{Serghiou
  et~al.}(1999{\natexlab{b}})\citenamefont{Serghiou, Miehe, Tschauner, Zerr,
  and Boehler}}]{r-ge32}
\bibinfo{author}{\bibfnamefont{G.}~\bibnamefont{Serghiou}},
  \bibinfo{author}{\bibfnamefont{G.}~\bibnamefont{Miehe}},
  \bibinfo{author}{\bibfnamefont{O.}~\bibnamefont{Tschauner}},
  \bibinfo{author}{\bibfnamefont{A.}~\bibnamefont{Zerr}}, \bibnamefont{and}
  \bibinfo{author}{\bibfnamefont{R.}~\bibnamefont{Boehler}},
  \bibinfo{journal}{J. Chem. Phys.} \textbf{\bibinfo{volume}{111}}
  (\bibinfo{year}{1999}{\natexlab{b}}).

\bibitem[{\citenamefont{Ching et~al.}(2001)\citenamefont{Ching, Mo, Tanaka, and
  Yoshiya}}]{PhysRevB.63.064102}
\bibinfo{author}{\bibfnamefont{W.~Y.} \bibnamefont{Ching}},
  \bibinfo{author}{\bibfnamefont{S.-D.} \bibnamefont{Mo}},
  \bibinfo{author}{\bibfnamefont{I.}~\bibnamefont{Tanaka}}, \bibnamefont{and}
  \bibinfo{author}{\bibfnamefont{M.}~\bibnamefont{Yoshiya}},
  \bibinfo{journal}{Phys. Rev. B} \textbf{\bibinfo{volume}{63}},
  \bibinfo{pages}{064102} (\bibinfo{year}{2001}).

\bibitem[{\citenamefont{Hu et~al.}(2006)\citenamefont{Hu, Cheng, Huang, Wu, and
  Xie}}]{c3n4_op}
\bibinfo{author}{\bibfnamefont{J.}~\bibnamefont{Hu}},
  \bibinfo{author}{\bibfnamefont{W.}~\bibnamefont{Cheng}},
  \bibinfo{author}{\bibfnamefont{S.}~\bibnamefont{Huang}},
  \bibinfo{author}{\bibfnamefont{D.}~\bibnamefont{Wu}}, \bibnamefont{and}
  \bibinfo{author}{\bibfnamefont{Z.}~\bibnamefont{Xie}},
  \bibinfo{journal}{Appl. Phys. Lett.} \textbf{\bibinfo{volume}{89}},
  \bibinfo{eid}{261117} (\bibinfo{year}{2006}).

\bibitem[{\citenamefont{Kohn and Sham}(1965)}]{KSE}
\bibinfo{author}{\bibfnamefont{W.}~\bibnamefont{Kohn}} \bibnamefont{and}
  \bibinfo{author}{\bibfnamefont{L.~J.} \bibnamefont{Sham}},
  \bibinfo{journal}{Phys. Rev.} \textbf{\bibinfo{volume}{140}},
  \bibinfo{pages}{A1133} (\bibinfo{year}{1965}).

\bibitem[{\citenamefont{Ching and Lin}(1975)}]{olcao}
\bibinfo{author}{\bibfnamefont{W.~Y.} \bibnamefont{Ching}} \bibnamefont{and}
  \bibinfo{author}{\bibfnamefont{C.~C.} \bibnamefont{Lin}},
  \bibinfo{journal}{Phys. Rev. B} \textbf{\bibinfo{volume}{12}},
  \bibinfo{pages}{5536} (\bibinfo{year}{1975}).

\bibitem[{\citenamefont{Hedin}(1965)}]{gw1}
\bibinfo{author}{\bibfnamefont{L.}~\bibnamefont{Hedin}},
  \bibinfo{journal}{Phys. Rev.} \textbf{\bibinfo{volume}{139}},
  \bibinfo{pages}{A796} (\bibinfo{year}{1965}).

\bibitem[{\citenamefont{Kresse et~al.}(2012)\citenamefont{Kresse, Marsman,
  Hintzsche, and Flage-Larsen}}]{PhysRevB.85.045205}
\bibinfo{author}{\bibfnamefont{G.}~\bibnamefont{Kresse}},
  \bibinfo{author}{\bibfnamefont{M.}~\bibnamefont{Marsman}},
  \bibinfo{author}{\bibfnamefont{L.~E.} \bibnamefont{Hintzsche}},
  \bibnamefont{and}
  \bibinfo{author}{\bibfnamefont{E.}~\bibnamefont{Flage-Larsen}},
  \bibinfo{journal}{Phys. Rev. B} \textbf{\bibinfo{volume}{85}},
  \bibinfo{pages}{045205} (\bibinfo{year}{2012}).

\bibitem[{\citenamefont{Gao et~al.}(2013)\citenamefont{Gao, Cai, and
  Xu}}]{Gao2013292}
\bibinfo{author}{\bibfnamefont{S.-P.} \bibnamefont{Gao}},
  \bibinfo{author}{\bibfnamefont{G.}~\bibnamefont{Cai}}, \bibnamefont{and}
  \bibinfo{author}{\bibfnamefont{Y.}~\bibnamefont{Xu}},
  \bibinfo{journal}{Comput. Mater. Sci.} \textbf{\bibinfo{volume}{67}},
  \bibinfo{pages}{292 } (\bibinfo{year}{2013}).

\bibitem[{\citenamefont{Xu and Gao}(2012)}]{Xu201211072}
\bibinfo{author}{\bibfnamefont{Y.}~\bibnamefont{Xu}} \bibnamefont{and}
  \bibinfo{author}{\bibfnamefont{S.-P.} \bibnamefont{Gao}},
  \bibinfo{journal}{Int. J. Hydrogen Energ.} \textbf{\bibinfo{volume}{37}},
  \bibinfo{pages}{11072 } (\bibinfo{year}{2012}).

\bibitem[{\citenamefont{Bl\"ochl}(1994)}]{PhysRevB.50.17953}
\bibinfo{author}{\bibfnamefont{P.~E.} \bibnamefont{Bl\"ochl}},
  \bibinfo{journal}{Phys. Rev. B} \textbf{\bibinfo{volume}{50}},
  \bibinfo{pages}{17953} (\bibinfo{year}{1994}).

\bibitem[{\citenamefont{G\'omez-Abal et~al.}(2008)\citenamefont{G\'omez-Abal,
  Li, Scheffler, and Ambrosch-Draxl}}]{PRL.101.106404}
\bibinfo{author}{\bibfnamefont{R.}~\bibnamefont{G\'omez-Abal}},
  \bibinfo{author}{\bibfnamefont{X.}~\bibnamefont{Li}},
  \bibinfo{author}{\bibfnamefont{M.}~\bibnamefont{Scheffler}},
  \bibnamefont{and}
  \bibinfo{author}{\bibfnamefont{C.}~\bibnamefont{Ambrosch-Draxl}},
  \bibinfo{journal}{Phys. Rev. Lett.} \textbf{\bibinfo{volume}{101}},
  \bibinfo{pages}{106404} (\bibinfo{year}{2008}).

\bibitem[{\citenamefont{Li et~al.}(2012)\citenamefont{Li, Gómez-Abal, Jiang,
  Ambrosch-Draxl, and Scheffler}}]{AE_GW}
\bibinfo{author}{\bibfnamefont{X.-Z.} \bibnamefont{Li}},
  \bibinfo{author}{\bibfnamefont{R.}~\bibnamefont{Gómez-Abal}},
  \bibinfo{author}{\bibfnamefont{H.}~\bibnamefont{Jiang}},
  \bibinfo{author}{\bibfnamefont{C.}~\bibnamefont{Ambrosch-Draxl}},
  \bibnamefont{and}
  \bibinfo{author}{\bibfnamefont{M.}~\bibnamefont{Scheffler}},
  \bibinfo{journal}{New J. Phys.} \textbf{\bibinfo{volume}{14}},
  \bibinfo{pages}{023006} (\bibinfo{year}{2012}).

\bibitem[{\citenamefont{Hybertsen and Louie}(1986)}]{gw2}
\bibinfo{author}{\bibfnamefont{M.~S.} \bibnamefont{Hybertsen}}
  \bibnamefont{and} \bibinfo{author}{\bibfnamefont{S.~G.} \bibnamefont{Louie}},
  \bibinfo{journal}{Phys. Rev. B} \textbf{\bibinfo{volume}{34}},
  \bibinfo{pages}{5390} (\bibinfo{year}{1986}).

\bibitem[{\citenamefont{Godby and Needs}(1989)}]{GNPPM}
\bibinfo{author}{\bibfnamefont{R.~W.} \bibnamefont{Godby}} \bibnamefont{and}
  \bibinfo{author}{\bibfnamefont{R.~J.} \bibnamefont{Needs}},
  \bibinfo{journal}{Phys. Rev. Lett.} \textbf{\bibinfo{volume}{62}},
  \bibinfo{pages}{1169} (\bibinfo{year}{1989}).

\bibitem[{\citenamefont{von~der Linden and Horsch}(1988)}]{vdLHPPM}
\bibinfo{author}{\bibfnamefont{W.}~\bibnamefont{von~der Linden}}
  \bibnamefont{and} \bibinfo{author}{\bibfnamefont{P.}~\bibnamefont{Horsch}},
  \bibinfo{journal}{Phys. Rev. B} \textbf{\bibinfo{volume}{37}},
  \bibinfo{pages}{8351} (\bibinfo{year}{1988}).

\bibitem[{\citenamefont{Engel and Farid}(1993)}]{EFPPM}
\bibinfo{author}{\bibfnamefont{G.~E.} \bibnamefont{Engel}} \bibnamefont{and}
  \bibinfo{author}{\bibfnamefont{B.}~\bibnamefont{Farid}},
  \bibinfo{journal}{Phys. Rev. B} \textbf{\bibinfo{volume}{47}},
  \bibinfo{pages}{15931} (\bibinfo{year}{1993}).

\bibitem[{\citenamefont{Shaltaf et~al.}(2008)\citenamefont{Shaltaf, Rignanese,
  Gonze, Giustino, and Pasquarello}}]{PhysRevLett.100.186401}
\bibinfo{author}{\bibfnamefont{R.}~\bibnamefont{Shaltaf}},
  \bibinfo{author}{\bibfnamefont{G.-M.} \bibnamefont{Rignanese}},
  \bibinfo{author}{\bibfnamefont{X.}~\bibnamefont{Gonze}},
  \bibinfo{author}{\bibfnamefont{F.}~\bibnamefont{Giustino}}, \bibnamefont{and}
  \bibinfo{author}{\bibfnamefont{A.}~\bibnamefont{Pasquarello}},
  \bibinfo{journal}{Phys. Rev. Lett.} \textbf{\bibinfo{volume}{100}},
  \bibinfo{pages}{186401} (\bibinfo{year}{2008}).

\bibitem[{\citenamefont{Stankovski et~al.}(2011)\citenamefont{Stankovski,
  Antonius, Waroquiers, Miglio, Dixit, Sankaran, Giantomassi, Gonze, C\^ot\'e,
  and Rignanese}}]{PhysRevB.84.241201}
\bibinfo{author}{\bibfnamefont{M.}~\bibnamefont{Stankovski}},
  \bibinfo{author}{\bibfnamefont{G.}~\bibnamefont{Antonius}},
  \bibinfo{author}{\bibfnamefont{D.}~\bibnamefont{Waroquiers}},
  \bibinfo{author}{\bibfnamefont{A.}~\bibnamefont{Miglio}},
  \bibinfo{author}{\bibfnamefont{H.}~\bibnamefont{Dixit}},
  \bibinfo{author}{\bibfnamefont{K.}~\bibnamefont{Sankaran}},
  \bibinfo{author}{\bibfnamefont{M.}~\bibnamefont{Giantomassi}},
  \bibinfo{author}{\bibfnamefont{X.}~\bibnamefont{Gonze}},
  \bibinfo{author}{\bibfnamefont{M.}~\bibnamefont{C\^ot\'e}}, \bibnamefont{and}
  \bibinfo{author}{\bibfnamefont{G.-M.} \bibnamefont{Rignanese}},
  \bibinfo{journal}{Phys. Rev. B} \textbf{\bibinfo{volume}{84}},
  \bibinfo{pages}{241201} (\bibinfo{year}{2011}).

\bibitem[{\citenamefont{Larson et~al.}(2013)\citenamefont{Larson, Dvorak, and
  Wu}}]{PhysRevB.88.125205}
\bibinfo{author}{\bibfnamefont{P.}~\bibnamefont{Larson}},
  \bibinfo{author}{\bibfnamefont{M.}~\bibnamefont{Dvorak}}, \bibnamefont{and}
  \bibinfo{author}{\bibfnamefont{Z.}~\bibnamefont{Wu}}, \bibinfo{journal}{Phys.
  Rev. B} \textbf{\bibinfo{volume}{88}}, \bibinfo{pages}{125205}
  (\bibinfo{year}{2013}).

\bibitem[{\citenamefont{Aryasetiawan and Gunnarsson}(1998)}]{gw_review}
\bibinfo{author}{\bibfnamefont{F.}~\bibnamefont{Aryasetiawan}}
  \bibnamefont{and}
  \bibinfo{author}{\bibfnamefont{O.}~\bibnamefont{Gunnarsson}},
  \bibinfo{journal}{Rep. Prog. Phys.} \textbf{\bibinfo{volume}{61}},
  \bibinfo{pages}{237} (\bibinfo{year}{1998}).

\bibitem[{\citenamefont{Kozhevnikov et~al.}(2010)\citenamefont{Kozhevnikov,
  Eguiluz, and Schulthess}}]{elk}
\bibinfo{author}{\bibfnamefont{A.}~\bibnamefont{Kozhevnikov}},
  \bibinfo{author}{\bibfnamefont{A.~G.} \bibnamefont{Eguiluz}},
  \bibnamefont{and} \bibinfo{author}{\bibfnamefont{T.~C.}
  \bibnamefont{Schulthess}}, in \emph{\bibinfo{booktitle}{SC'10 Proceedings of
  the 2010 ACM/IEEE International Conference for High Performance Computing,
  Networking, Storage, and Analysis (IEEE Computer Society, Washington, DC,
  2010)}} (\bibinfo{year}{2010}), pp. \bibinfo{pages}{1--10}.

\bibitem[{\citenamefont{Ku and Eguiluz}(2002)}]{Ku_GW}
\bibinfo{author}{\bibfnamefont{W.}~\bibnamefont{Ku}} \bibnamefont{and}
  \bibinfo{author}{\bibfnamefont{A.~G.} \bibnamefont{Eguiluz}},
  \bibinfo{journal}{Phys. Rev. Lett.} \textbf{\bibinfo{volume}{89}},
  \bibinfo{pages}{126401} (\bibinfo{year}{2002}).

\bibitem[{\citenamefont{Usuda et~al.}(2002)\citenamefont{Usuda, Hamada, Kotani,
  and van Schilfgaarde}}]{PhysRevB.66.125101}
\bibinfo{author}{\bibfnamefont{M.}~\bibnamefont{Usuda}},
  \bibinfo{author}{\bibfnamefont{N.}~\bibnamefont{Hamada}},
  \bibinfo{author}{\bibfnamefont{T.}~\bibnamefont{Kotani}}, \bibnamefont{and}
  \bibinfo{author}{\bibfnamefont{M.}~\bibnamefont{van Schilfgaarde}},
  \bibinfo{journal}{Phys. Rev. B} \textbf{\bibinfo{volume}{66}},
  \bibinfo{pages}{125101} (\bibinfo{year}{2002}).

\bibitem[{\citenamefont{He et~al.}(2001)\citenamefont{He, Sekine, Kobayashi,
  and Kimoto}}]{r-ge33}
\bibinfo{author}{\bibfnamefont{H.}~\bibnamefont{He}},
  \bibinfo{author}{\bibfnamefont{T.}~\bibnamefont{Sekine}},
  \bibinfo{author}{\bibfnamefont{T.}~\bibnamefont{Kobayashi}},
  \bibnamefont{and} \bibinfo{author}{\bibfnamefont{K.}~\bibnamefont{Kimoto}},
  \bibinfo{journal}{J. Appl. Phys.} \textbf{\bibinfo{volume}{90}}
  (\bibinfo{year}{2001}).

\bibitem[{\citenamefont{Boyko et~al.}(2010)\citenamefont{Boyko, Bailey, Moewes,
  and McMillan}}]{PhysRevB.81.155207}
\bibinfo{author}{\bibfnamefont{T.~D.} \bibnamefont{Boyko}},
  \bibinfo{author}{\bibfnamefont{E.}~\bibnamefont{Bailey}},
  \bibinfo{author}{\bibfnamefont{A.}~\bibnamefont{Moewes}}, \bibnamefont{and}
  \bibinfo{author}{\bibfnamefont{P.~F.} \bibnamefont{McMillan}},
  \bibinfo{journal}{Phys. Rev. B} \textbf{\bibinfo{volume}{81}},
  \bibinfo{pages}{155207} (\bibinfo{year}{2010}).

\bibitem[{\citenamefont{Hoffman et~al.}(1995)\citenamefont{Hoffman,
  Prakash~Rangarajan, Athavale, Economou, Liu, Zheng, and Chu}}]{bge2}
\bibinfo{author}{\bibfnamefont{D.~M.} \bibnamefont{Hoffman}},
  \bibinfo{author}{\bibfnamefont{S.}~\bibnamefont{Prakash~Rangarajan}},
  \bibinfo{author}{\bibfnamefont{S.~D.} \bibnamefont{Athavale}},
  \bibinfo{author}{\bibfnamefont{D.~J.} \bibnamefont{Economou}},
  \bibinfo{author}{\bibfnamefont{J.-R.} \bibnamefont{Liu}},
  \bibinfo{author}{\bibfnamefont{Z.}~\bibnamefont{Zheng}}, \bibnamefont{and}
  \bibinfo{author}{\bibfnamefont{W.-K.} \bibnamefont{Chu}},
  \bibinfo{journal}{J. Vac. Sci. Technol. A} \textbf{\bibinfo{volume}{13}}
  (\bibinfo{year}{1995}).

\end{thebibliography}

\end{document}